\documentclass[aps,pra,twocolumn,showpacs]{revtex4-1}
\usepackage{times,amsmath,amssymb,amsfonts,graphicx,color}

\newcommand{\ket}[1]{\vert #1 \rangle} %
\newcommand{\bra}[1]{\langle #1 \vert} %

\newcommand{\id}{\mathbb{I}}
\newcommand{\expected}[1]{\mathbb{E}\left[#1\right]}
\begin{document}
\title{Engineering decoherence for two-qubit systems interacting
with a classical environment}
\author{Matteo A. C. Rossi}
\affiliation{Dipartimento di Fisica e Scienze della 
Terra ``Macedonio Melloni'',Universit\`a 
degli Studi di Parma, Parco Area delle Scienze 7/A, 
I-43124 Parma, Italy\email{matteo.rossi11@studenti.unipr.it}}
\author{Claudia Benedetti}
\affiliation{Dipartimento di Fisica, Universit\`a degli 
Studi di Milano, I-20133 Milano, Italy} %
\author{Matteo G. A. Paris}
\affiliation{Dipartimento di Fisica, Universit\`a 
degli Studi di Milano, I-20133 Milano, Italy} %
\date{\today}
%%%%%%%%%%%%%%%%%%%%%%%%%%%%%%%%%%%%%%%%%%%%%%%%%%%%%%%%%%%%%%%%%%%
\begin{abstract}
We address the dynamics of a two-qubit system interacting with a
classical dephasing environment driven by a Gaussian stochastic process.
Upon introducing the concept of {\em entanglement-preserving time}, we
compare the degrading effects of different environments, e.g. those
described by Ornstein-Uhlenbeck or fractional noise.  In particular, we
consider pure Bell states and mixtures of Bell states and study the
typical values of the entanglement-preserving time for both independent and
common environments. We found that engineering environments towards
fractional Gaussian noise is useful to preserve entanglement as well as
to improve its robustness against noise. We also address entanglement
sudden death by studying the {\em entanglement-survival time} as a
function of the initial negativity. We found that: i) the survival time
is bounded from below by an increasing function of the initial
negativity, ii) the survival time depends only slightly on the process
used to describe the environment and exhibits typicality.
Overall, our results show that engineering the environment has only
a slight influence over the entanglement-survival time, i.e. the occurence
of entanglement sudden-death, while it represents a valuable resource 
to increase the entanglement-preserving time, i.e. to maintain entanglement 
closer to the initial level for a longer interaction time.
\end{abstract}
\maketitle
%%%% 
\section{Introduction}
The unavoidable interaction of a quantum system with its environment
generally causes decoherence and a loss of quantumness. On the other
hand, the possibility to perform quantum operations within the coherence
time of a quantum system lies at the heart of quantum information
processing. A deep understanding of the decoherence mechanisms in
quantum systems, together with the capability to engineer the
environment in order to reduce its detrimental effects, are thus
essential steps toward the development of quantum technologies.
\par
The interaction of a quantum system with its environment may be
described using either a classical or a quantum mechanical picture of
the environment. Understanding whether and in which conditions the two
descriptions are equivalent is still a debated topic
\cite{helm09,helm11,crow2014classical,wayne13}.  When the environment
has many degrees of freedom and/or a structured noise spectrum, a
quantum description may be challenging, and the approximations may be
crude enough to prevent a reliable description of the dynamics.  In
these situations, a classical description may be convenient and also
more accurate. Several systems of interest belong to these categories and
many efforts have been devoted to study situations where quantum systems
are affected by classical noise.  Examples include the dynamics of
quantum correlations
\cite{yu10,arr10,li11,benedetti12,bordone12,rlf12,rlf13,xu13,arr14}, 
the simulation of motional averaging \cite{paraoanu14}, or decoherence in
solid state qubits
\cite{meno,tsai,pal0x,bukard,bergli12,pal12,nonmark,benedetti2013dynamics,
mannone12,rev14,paris14} and the characterization of the
environment using quantum probes \cite{benedetti14,parisF14}.  When the
environment affecting the quantum system may be described as collection
of fluctuators, a Gaussian statistics for the noise can be assumed
\cite{bergli09,astafiev2004quantum}. Moreover, the Gaussian
approximation is valid even in the presence of non-Gaussian noise, as
far as the coupling with the environment is weak \cite{bergli06,abel08}.
\par
In this paper, we address the dynamics of entanglement for a two-qubit
system subject to a classical noise induced by a Gaussian stochastic
process. Specifically, we consider the case where the typical
frequencies of the system are larger compared to those of the
environment, so that the system dynamics can be described as a pure
dephasing \cite{averin04,shnirma,shibata,sarma08,sarma13b,rev14}. 
Dephasing induced by classical noise has been studied
previously \cite{yu2006sudden,yu10}, and it is known to induce a
monotonic decay of entanglement, including the phenomenon of sudden
death \cite{yu2009sudden} (ESD) i.e. the  transition from an entangled to a
separable state after a finite interaction time.  Here we quantitatively
compare the degrading effects of different kinds of environments by
defining the {\em entanglement-preserving time} and the {\em
entanglement-survival time} and by studying their dependence on the nature
and on the parameters of the stochastic process that models the
environment.  We focus on two paradigmatic examples of Gaussian
processes describing normal and anomalous diffusion processes: the
Ornstein-Uhlenbeck process \cite{palma03,spagnolo09,paris14} and the
fractional Gaussian noise \cite{mandelbrot1968fractional}. 
\par
This paper is organized as follows: in Sec. \ref{sec:model} we 
describe the physical model that accounts for the system-environment 
interaction and introduce the Gaussian processes that drive 
the noise. In Sec. \ref{sec:results} we look at the dynamics of the 
system and analyze in some detail the dependence of the 
entanglement-preserving time and the entanglement-survival time on the 
nature of the Gaussian process and the initial state of the system.
Sec. \ref{sec:conclusions} closes the paper with some concluding 
remarks.
%%%
\section{The physical model}
\label{sec:model}
We consider a system of two non-interacting, identical qubits,
characterized by the same energy splitting $\omega_0$ and coupled to two
external classical fluctuating fields.  The effective Hamiltonian is
thus of the form \begin{equation}\label{eq:Hamiltonian}
\mathcal{H}(t) =  \mathcal{H}_1(t) 
\otimes \id_2 + \id_1 \otimes \mathcal{H}_2(t),
\end{equation}
where the local Hamiltonians are
\begin{equation}\label{eq:local_Hamiltonian}
\mathcal{H}_i (t)= [\omega_0 + \lambda B_i(t)]\sigma_z.
\end{equation}
Here, $\lambda$ is a coupling constant and $B_i(t)$ is an external
classical field acting on each qubit, which we describe by means of a
zero-mean Gaussian stochastic process. We consider both the case in
which the two qubits are interacting with two independent environments,
i.e.  $B_1(t)$ and $B_2(t)$ are totally uncorrelated, and the case in
which the two qubits are subject to a common environment,  $B_1(t) =
B_2(t)$.
\par
The Hamiltonian \eqref{eq:Hamiltonian} models an effective interaction 
between a quantum system and a noisy environment having characteristic 
frequencies  much smaller than the typical frequencies of the system 
$\omega_0$.  The Hamiltonian in Eqs. \eqref{eq:Hamiltonian} and 
\eqref{eq:local_Hamiltonian} can also describe a two-level quantum degree 
of freedom coupled to a classical degree of freedom, for example the spin 
of a spin-$\frac 12$ particle undergoing a diffusion process in an external 
field.
\par
A Gaussian process can be described completely by its second order 
statistics, i.e. by its mean $\mu$  and its autocorrelation function 
$K$, in formula:
\begin{align}	
\mu(t) &= \expected{B(t)} = 0 \\
K(t,t') &= \expected{B(t)B(t')}
\end{align}
where $\expected{\cdot}$ denotes the average over all possible 
realizations of the process $B(t)$. The characteristic function 
of a Gaussian  process is defined as \cite{puri2001mathematical}
\begin{align}\label{eq:characteristic_function_Gaussian}
&\expected{\exp\left(i \int_0^t\!\! ds\, f(s) B(s)\right)}  = \notag \\ 
&\quad \exp\left[- \frac 12 \int_0^t\int_0^t\!\! ds\, ds'\, 
f(s) K(s,s') f(s') \right],
\end{align}
where $f(t)$ is an arbitrary function of time. If $f = \kappa$ is 
constant with respect to time, Eq. 
\eqref{eq:characteristic_function_Gaussian} rewrites as
\begin{equation}\label{eq:characteristic_function_2}
\expected{\exp\left(\pm i \kappa \int_0^t\!\!ds\,  B(s)\right)} 
= \exp \left[- \frac12 \kappa^2 \beta(t) \right]
\end{equation}
where
\begin{equation}
\beta(t) = \int_0^t\int_0^t\!\! ds\,ds'\, K(s,s').\label{beta}
\end{equation}
\par
In this work, we focus on two paradigmatic Gaussian processes: the
Ornstein-Uhlenbeck (OU) process and the fractional Gaussian noise (fGn).
The OU process describes a diffusion process with friction and it is
characterized by the autocorrelation function 
\begin{equation}
K_{\text{OU}}(t-t') = \frac \gamma 2 \exp(-\gamma |t-t'|),
\end{equation} 
where $\gamma = \tau^{-1}$ plays the role of a memory parameter 
and $\tau$ is the correlation time of the process. 
For increasing $\gamma$ the noise spectrum becomes
broader and in the limit $\gamma\gg1$ one achieves white noise.
The fractional Gaussian noise describes anomalous diffusion processes,
with a diffusion coefficient proportional to $t^{2H}$, where $H\in
(0,1)$ is known as the Hurst parameter. The covariance function may 
be written 
\begin{equation}
K_{\text{fGn}}(t-t') = \frac 12 (|t|^{2H}+ |t'|^{2H}- |t-t'|^{2H}).
\end{equation}
When $H=1/2$ we have $K_{\text{fGn}}(t-t') = 
\min (t,t')$ and the fGn reduces to
the Wiener process (i.e. Brownian motion). When $H > \frac 12$, the
increments of the process have positive correlation and the regime is
called super-diffusive; when $H<\frac 12$, we are in the sub-diffusive
regime and the increments are negatively correlated.
\par
The $\beta$ functions \eqref{beta} for 
the OU and fGn processes are given by:
\begin{align}
\beta_{\text{OU}}(t) &= \frac 1 \gamma (e^{-\gamma t} + \gamma t - 1 ) \\
\beta_{\text{fGn}}(t) &= \frac{t^{2H+2}}{2H+2}.
\end{align}
The evolution operator $U(t)$ for a given realization of the process
$B_i(t)$, is expressed as: 
\begin{align}
U(t)  =& \exp \left[-i \int_0^t \mathcal{H}(s)ds\right] = \notag \\
 =& \exp\{-i[\omega_0 t + \lambda \varphi_1(t)]\sigma_z \}\notag \\ 
 & \otimes 
\exp\{-i[\omega_0t+\lambda \varphi_2(t)]\sigma_z\}
\end{align}
where we defined the phase noise $\varphi_i(t) = \int_0^t\! ds\, B_i(s)$. 
If the system is initially prepared in the state $\rho_0$, the density
matrix at a time $t$ is given by the expected value of the evolved
density matrix over all possible realizations of the stochastic
processes, i.e. 
\begin{equation}\label{eq:rho_t}
\rho(t) = \expected{U(t)\rho_0 U^\dagger(t)}.
\end{equation}
As initial state, we consider a system prepared in a Bell-state mixture:
\begin{align}
\rho_0 =& c_1 \ket{\Phi^+}\bra{\Phi^+} + c_2 
\ket{\Phi^-}\bra{\Phi^-} \notag \\ &+ c_3 \ket{\Psi^+}
\bra{\Psi^+} + c_4 \ket{\Psi^-}\bra{\Psi^-} 
= \notag \\
& =\frac 14 \left( \id + \sum_{i=1}^3 
a_i \sigma_i \otimes \sigma_i \right), 
\label{eq:Bell_state_mixture}
\end{align} 
where $\ket{\Phi^{\pm}}=\frac{1}{\sqrt{2}}(\ket{00}
\pm\ket{11})$,  $\ket{\Psi^{\pm}}=\frac{1}{\sqrt{2}}
(\ket{01}\pm\ket{10})$,
and the $\sigma_i$ are the three Pauli matrices. 
The coefficients $c_i$ satisfy the condition $\sum c_i = 1$, and 
are related to the $a_i$ through the equalities:
\begin{align}
 a_1&=c_1-c_2+c_3-c_4\nonumber\\
 a_2&=-c_1+c_2+c_3-c_4\\
 a_3&=c_1+c_2-c_3-c_4\nonumber
\end{align}
We evaluate the entanglement by means of the 
negativity 
\begin{equation}
N(\rho) = 2 \left|\sum_i \lambda^-_i \right|,
\end{equation}
where $\lambda_i^-$ are the negative eigenvalues of the partial
transpose of the system density matrix.  Negativity  is
zero for separable states and one for maximally entangled states, such
as pure Bell states.
%%%%%%%%%%%%%%%%%%%%%%%%%%%%%%%%%%
\section{Results}
\label{sec:results}
\begin{figure}[h!]
\centering
\includegraphics[width=0.48\columnwidth]{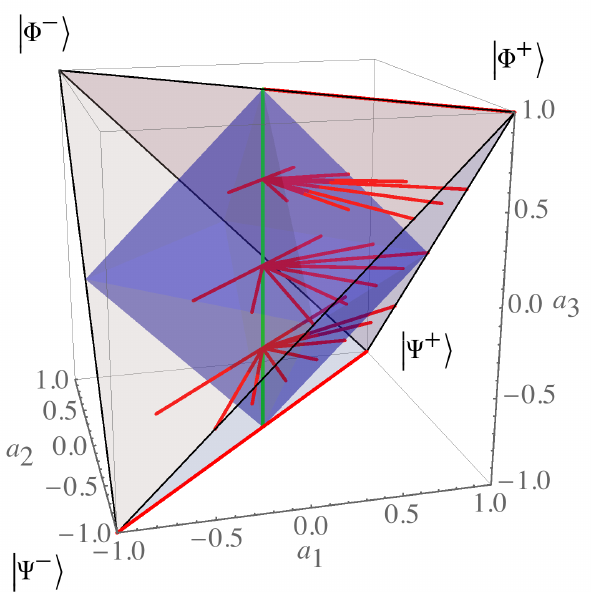} 
\includegraphics[width=0.48\columnwidth]{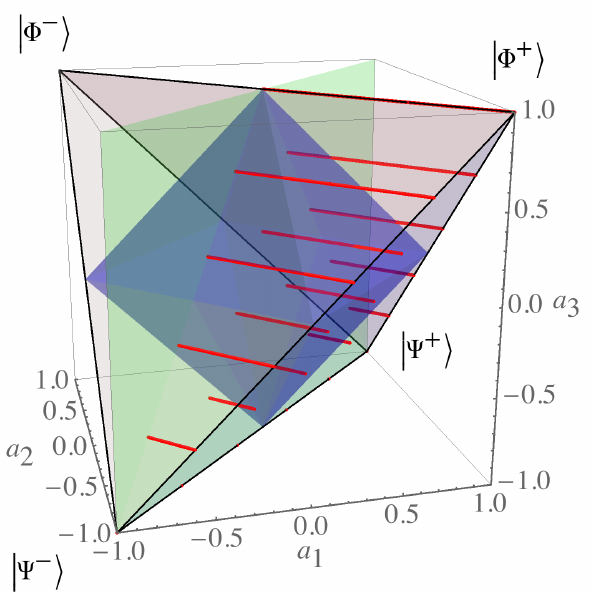} 
\caption{Trajectories of the system in the space of parameters
$\{a_1,a_2,a_3\}$, for two independent environments (left) and for a
common environment (right). The Bell-state mixtures, Eq.
\eqref{eq:Bell_state_mixture}, form a tetrahedron. The set of separable
states is the dark-blue octahedron. The initial states are Bell-state
mixtures that lie on the surface of the tetrahedron. For independent
environments, the trajectories converge to the green line $a_1=a_2=0$. 
For a common environment, the trajectories are directed orthogonally to
the plane $a_1 = a_2$, shown in green. In both cases, $a_3$ remains
constant.} \label{fig:trajectories} \end{figure}

\begin{widetext}
\subsection{Independent environments}
\label{sub:independent_environments}
Here, we consider the case of independent environments, i.e. each qubit
is coupled to its own environment, described by the stochastic field 
$B_i(t)$.  In order to obtain the evolved density matrix of the 
system, we calculate the expectation
value in Eq. \eqref{eq:rho_t} over all possible realizations of the two
uncorrelated  processes $B_1(t)$ and $B_2(t)$. The evolved density
matrix for the two qubits can be written explicitly by using Eq.
\eqref{eq:characteristic_function_2}. We find
\begin{equation}\label{eq:mean_rho_se}
%\resizebox{.91\textwidth}{!}{$\displaystyle
\rho(t) = \frac 12 \left(
\begin{array}{cccc}
 \left(c_1+c_2\right) & 0 & 0 &  
 e^{-4\lambda^2 \beta - 4 i \omega_0 t} \left(c_1-c_2\right) \\ 0 &  
 \left(c_3+c_4\right) &  e^{-4 \lambda^2 \beta } 
 \left(c_3-c_4\right) & 0 \\ 0 &  e^{-4 \lambda^2 
 \beta } \left(c_3-c_4\right) &  \left(c_3+c_4\right) & 0 \\
  e^{-4 \lambda^2 \beta +4 i \omega_0 t} 
  \left(c_1-c_2\right) & 0 & 0 &  \left(c_1+c_2\right) \\
\end{array}
\right)%,$}
\end{equation}
that is, a pure dephasing map.
By applying the local unitary transformation 
$e^{i \omega_0 t \sigma_z} \otimes e^{i \omega_0 t 
\sigma_z}$, we can write $\rho(t)$ in the diagonal Bloch form
\begin{equation}
\rho(t) = \frac 14 (\id + e^{-4\lambda^2 \beta(t)}a_1 
\sigma_x \otimes \sigma_x + e^{-4\lambda^2 \beta(t)}
a_2 \sigma_y \otimes \sigma_y + a_3 \sigma_z \otimes \sigma_z),
	\label{rhot2}
\end{equation}
where $a_1,a_2$ and $a_3$ are the components of the initial state
$\rho_0$.  Since the density matrix \eqref{rhot2} depends on time only
through the function $\beta(t)$, the system will  reach the separable
steady state $\rho(t) = \frac 14 (\id + a_3 \sigma_z \otimes \sigma_z)$
for $t\rightarrow \infty$.  The trajectories of the evolved states in
the $a_i$-parameter space are shown in Fig.~\ref{fig:trajectories}
(left).  We notice that, with the exception of  initial  Bell states,
the trajectories of the system actually enter the set of separable
states at a finite time, thus showing a sudden death of entanglement.
\par
The negativity as a function of time, for an initial arbitrary
Bell-state mixture, is given by: 
\begin{align} 
N(t)  = & \frac{1}{2} \left(\left| c_1+c_2 + 
e^{-4\lambda^2\beta (t)} (c_3-c_4)\right| + 
\left|c_1+c_2 - e^{-4\lambda^2\beta (t)} 
(c_3-c_4)\right| + \right.  \notag \\
& \left. + \left| e^{-4\lambda^2\beta (t)} 
(c_1-c_2)+ c_3+c_4\right| +\left| -e^{-4
\lambda^2\beta (t)} (c_1-c_2)+c_3+c_4\right| 
\right)-1.		\label{eq:se_negativity}
\end{align}
As we can see from Eq. \eqref{eq:se_negativity}, the evolution of
negativity doesn't depend on the energy splitting $\omega_0$ of the two qubits.
%%%%
\subsection{Common environment} 
\label{sub:common_environment}
If the two qubits interact with the same environment, we 
can assume that $B_1(t) =B_2(t) = B(t)$ and thus
\begin{equation}
U(t) = \exp\{-i[\omega_0t + \lambda \varphi(t)]
\sigma_z \} \otimes \exp\{-i[\omega_0t+\lambda \varphi(t)]\sigma_z\}.
\end{equation}
The evolved density matrix at time $t$ is given by
\begin{equation}
%\resizebox{.9\textwidth}{!}{$\displaystyle
\rho(t) =\frac{1}{2}\left(
\begin{array}{cccc}
 \left(c_1+c_2\right) & 0 & 0 &  
 e^{-8 \lambda^2\beta -4 i \omega_0 t} \left(c_1-c_2\right) \\
 0 &  \left(c_3+c_4\right) &  \left(c_3-c_4\right) & 0 \\
 0 &  \left(c_3-c_4\right) &  \left(c_3+c_4\right) & 0 \\
 e^{- 8 \lambda^2	\beta + 4 i \omega_0 t} 
 \left(c_1-c_2\right) & 0 & 0 &  \left(c_1+c_2\right) \\
		\end{array}
		\right)%$
		%}
\end{equation}
and the Bloch-diagonal form of the state 
(after a local unitary transformation  
$e^{i \omega_0 t \sigma_z} \otimes e^{i \omega_0 t \sigma_z}$) is 
\begin{align}
\rho(t) = & \frac 14\left\{\id +  
\frac{1}{2} \left[e^{-8 \lambda^2 \beta(t)} 
(a_1-a_2)+a_1+a_2\right] \sigma_x \otimes \sigma_x + \notag \right. \\
& \left. + \frac{1}{2} \left[e^{-8 \lambda^2 \beta(t)} 
(a_2-a_1)+a_1+a_2\right] \sigma_y \otimes \sigma_y + 
a_3 \sigma_z \otimes \sigma_z \right \}.
\end{align}
In this case, the negativity as a function of time for an initial arbitrary mixture of Bell states, is 
\begin{align}	
N(t) = & \frac{1}{2} 
\left[ \left| e^{-8 \lambda^2 \beta (t)} (c_1-c_2)+
c_3+c_4\right| +\left| e^{-8 \lambda^2 \beta (t)} 
(c_2-c_1)+c_3+c_4\right| + \right. \notag \\
& \left.\vphantom{e^{-8 \lambda^2 \beta(t)}} + 
\left| 1-2 c_3\right| +\left| 1-2 c_4
\right| -2\right] \label{eq:negativity_ce}
\end{align}
The trajectories in the Bell-state tetrahedron are shown 
in Fig.~\ref{fig:trajectories} (right). They run orthogonally 
to the plane $a_1 = a_2$. By looking at the figure, we notice 
that the system experiences ESD when the initial state has  
$a_3 > 0$, except for mixtures of $\ket{\Phi^+}$ and $\ket{\Phi^-}$, 
for which $N(t)\rightarrow 0$ only for $t\rightarrow \infty$. 
For those Bell-state mixtures that are entangled and for which $a_3 <
0$, the trajectory runs parallel to the surface of the octahedron and
hence negativity is constant over time. This set also includes the two
Bell states $\ket{\Psi^\pm}$ which are stable states for the dephasing
dynamics.
\end{widetext}
$ $ 
%%%
%%%%
\subsection{Entanglement-preserving time}
\label{sec:entanglement_preserving_time}
\begin{figure}[h!]
\centering 
\includegraphics[width=.49\columnwidth]{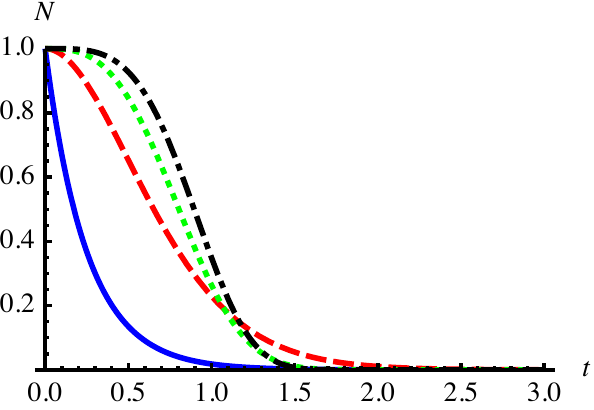}
\includegraphics[width=.49\columnwidth]{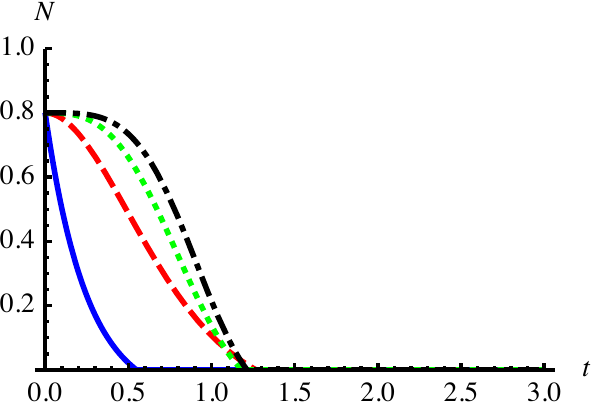}
\caption{Negativity as a function of the interaction time for 
an initially pure Bell state (left) and for the mixture 
$\rho= \frac {1}{10}\ket{\Phi^+}\bra{\Phi^+} + \frac{9}{10}
\ket{\Psi^+}\bra{\Psi^+}$ (right) interacting with
independent environments driven by different stochastic
processes: white noise (solid blue), OU with $\gamma = 1$ 
(red dashed), 
Wiener (green dotted), fGn with $H = 0.9$ (dot-dashed black).
For pure Bell states, the negativity
decreases smoothly to zero, while for mixtures of Bell states
ESD appears.} \label{fig:neg_vs_t} 
\end{figure}

The effect of the longitudinal field is to induce decoherence in the
form of a dephasing.  The entanglement,  computed by the negativity,
decays monotonically in time, as shown in Fig. \ref{fig:neg_vs_t}.  In
particular, depending on the initial state different behaviors of
quantum correlations appear: for initial Bell states, the negativity
goes asymptotically to zero, as a smooth function of time; on the
contrary, if the initial state is a mixture of Bell states, entanglement
displays sudden death, reaching zero abruptly. For a fixed initial state,
the robustness of quantum correlations depends on the nature of the
considered stochastic process: different expressions of the $\beta$
function give different decaying velocities for entanglement. 
%%%
\par
%%%
We now investigate the role of the different considered processes in
enhancing the system's ability to retain its coherence. To be
quantitative, we define the \emph {entanglement-preserving time} $t^*$
as the time at which the negativity of the system falls below a certain
threshold, that we fix at the ratio $r = 99\%$ of the initial negativity.
We first consider the case in which the initial state is a Bell state.
In this case, the negativity as a function of time is easily found to be
\begin{align}
	N_{\text{se}}(t) &= \exp [-4 \lambda^2 \beta(t)] \\
	N_{\text{ce}}(t) &= \exp [-8 \lambda^2 \beta(t)]
\end{align}
for the independent-environment and common-environment case, respectively.
Upon introducing the quantity $\beta^* = - 1/4 \log(r) \simeq 0.0025$, 
we may write the entanglement-preserving time as in
Table~\ref{tab:tstar}, where we show the dependencies of $t^*$ on the 
parameters of the processes, i.e. the inverse of the correlation time 
$\gamma$ for the Ornstein-Uhlenbeck process and the Hurst parameter 
$H$ for the fractional noise. We also report the results for white noise
(i.e. OU for $\gamma \rightarrow\infty$) and the Wiener process (i.e. fGn
with $H=\frac12$).
%%%
\begin{table}[h!]
\caption{The entanglement-preserving time $t^*$ for different environments 
and for an initial pure Bell state. The quantity $\beta^*$ is given 
by $\beta^* = - 1/4 \log(r) \simeq 0.0025$ and $W(z)$ is the Lambert
function, i.e. the principal solution of $z = W \exp W$.} 
{\begin{tabular}{ccc}
\toprule
Process & & $t^*$  \\ \hline 
& & \\
Ornstein-Uhlenbeck
&$\qquad$& $\frac1{\gamma} \left[\gamma\beta ^*+W\left(-e^{-\gamma\beta ^*-1}
\right)+1\right]$\\
White noise && $\beta^*$ \\
& & \\
fractional Gaussian noise 
&& $\left[(2 H + 2)\beta^*\right]^{\frac{1}{2 H+2}}$\\
Wiener && $[3 \beta^*]^{1/3}$ \\ 
\toprule
\end{tabular}}
\label{tab:tstar}
\end{table}
\par
\begin{figure}[h!]
\centering	
\includegraphics[width=0.48\columnwidth]{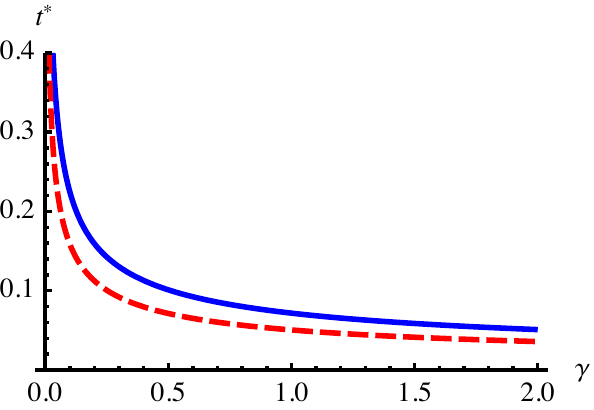} 
\includegraphics[width=0.48\columnwidth]{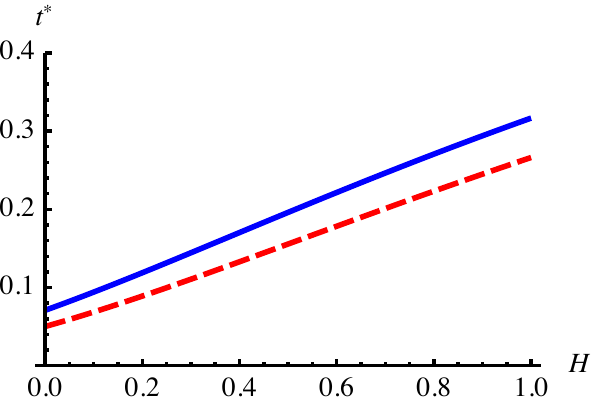}
\caption{The entanglement-preserving time $t^*$ as a function
of the characteristic parameter of the external field. We show 
results for Ornstein-Uhlenbeck (left) and fractional Gaussian 
noise (right) and for the case of independent (solid blue) and 
common (red dashed) environments.} 
\label{fig:tstar_OU_fBm}
\end{figure}

%%%
The entanglement-preserving time for OU and fGn is shown in 
Fig.~\ref{fig:tstar_OU_fBm} as a function of the characteristic
parameters $\gamma$ and $H$.  For the Ornstein-Uhlenbeck
process, in the limit of a quasi-static field, i.e. $\gamma\rightarrow
0$, the entanglement-preserving time diverges, $t^*\rightarrow
\infty$, such that system retains its coherence 
indefinitely, while in the
Markovian limit, $\gamma \rightarrow \infty$, $t^* \rightarrow \beta^*$,
recovering the behavior typical of the white noise.  In the case of fGn,
the dependence of $t^*$ on $H$ is well approximated by a linear relation
and the higher the diffusion coefficient, the longer the
entanglement-preserving time.  We also notice that, for vanishing $H$,
$t^*$ is comparable to the OU process with $\gamma = 1$.  Indeed,
we have that  $\beta_{\text{OU}}(t) \simeq \frac12 \gamma t^2$ for
small $t$ and $\beta_{\text{fGn}}(t) \simeq \frac12 t^2$ for vanishing $H$.
For general mixtures of Bell states, $t^*$ is always smaller than  the
case of pure  Bell states.  
%%%%%
%%%%%
\par
\begin{figure}[h!]	
\centering
\includegraphics[width=.48\columnwidth]{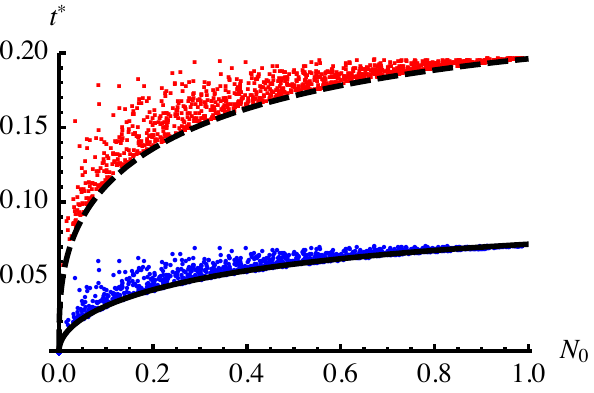} 
\includegraphics[width=.48\columnwidth]{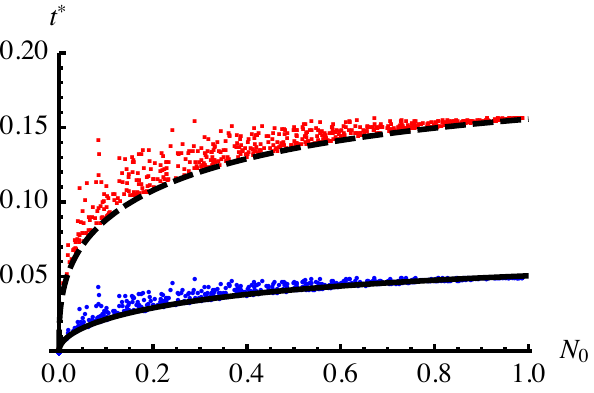}
\caption{The entanglement-preserving time $t^*$ 
(for a ratio $r=0.99$ to the initial negativity)
as a function of the initial negativity $N_0$ for randomly chosen
initial Bell-state mixtures. We show results for the Ornstein-Uhlenbeck
process with $\gamma = 1$ (blue points) and the Wiener process, i.e.
fractional Gaussian noise with $H = 1/2$ (red points). 
The solid and dashed black lines are the lower bounds for 
$t^*$ for the OU and Wiener process respectively, obtained 
from Eq.~\ref{eq:tstar_lower_bound}. Left: independent 
environments. Right: common environment.}
\label{fig:tstar_vs_neg}
\end{figure}

In Fig. \ref{fig:tstar_vs_neg} we show $t^*$ as a function of the
initial negativity $N_0$ for a set of randomly generated initial
Bell-mixed states interacting with OU and fGn external fields (blue and
red points respectively) either independently (left panel) or as a
common environment (right panel). As it is apparent from the plots, the
larger is the initial entanglement, the longer is the preserving time.
This is true both in the case of independent and common environments.
In the former case, the entanglement-preserving time is longer than in
case of a common bath, for a fixed value of the initial negativity.  In
both scenarios, the entanglement is more robust in the case of fGn,
rather than the OU process, with longer values of the preserving time
$t^*$.
\par
By looking at Fig. \ref{fig:tstar_vs_neg} we see that the values of
$t^*$ are not much dispersed. Rather, they concentrate around typical
values which strongly depend on the kind of environment and only
slightly on the initial negativity itself. Besides, the value of $t^*$
is bounded from below by an increasing function of the initial
negativity, the analytical expression of which can be obtained by
determining the entanglement-preserving time for mixtures of a $\Phi$
and a $\Psi$ Bell state. In this case, for a given ratio $r$ to 
the initial negativity, $t^*$ satisfies the equation
\begin{equation}
\label{eq:tstar_lower_bound}
\beta(t^*) = \frac{1}{4A} \log \left[\frac{N_0+1}{N_0 (2 r-1)+1}\right].
\end{equation}
where $A=1$ for independent environments and $A=2$ for a 
common environment. From Eq. (\ref{eq:tstar_lower_bound}) we 
obtain lower bounds to $t^*$ as a function of $N_0$, which
are shown (solid and dashed black lines) 
in Fig.~\ref{fig:tstar_vs_neg}.
%%%%
%%%%%%
\subsection{Entanglement survival time}
\label{sec:esd_time}
As previously discussed, the interaction of the two-qubit system with
the external classical field induces a sudden death of entanglement for
most of the Bell-state mixtures. In this section we study how the nature
of the stochastic Gaussian process affects the \emph{entanglement
survival time}, $t_{\text{\tiny ES}}$, i.e. the time at which the state
becomes separable and its negativity goes to zero.
\par
\begin{figure}[h!]
\centering
\includegraphics[width=.48\columnwidth]{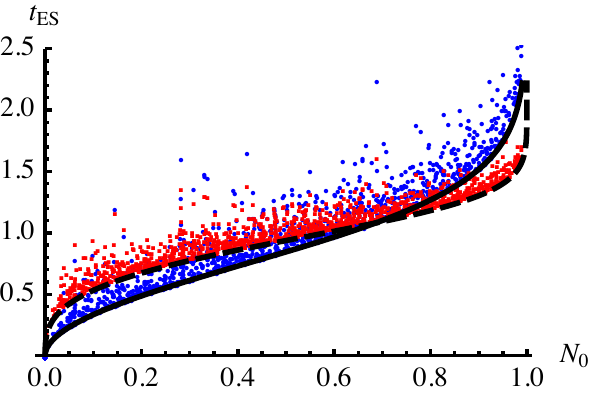} 
\includegraphics[width=.48\columnwidth]{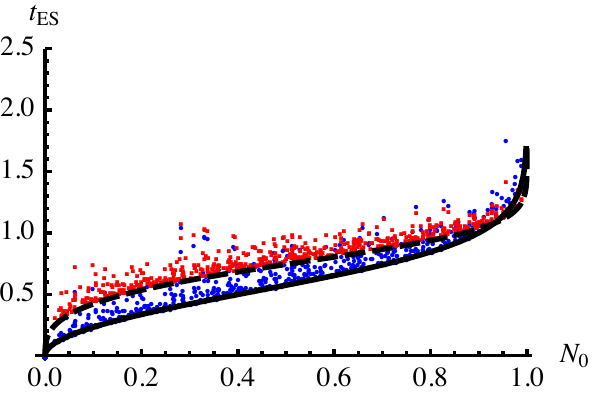}
\caption{The entanglement-survival time $t_{\text{\tiny ES}}$ 
as a function of the initial negativity $N_0$ for randomly 
chosen (initial) Bell-state mixtures, for the Ornstein-Uhlenbeck 
process with $\gamma = 1$ (blue)
and the Wiener process, i.e. fractional Gaussian noise with $H = 1/2$
(red). Left: independent environments. The solid and dashed lines are
the lower bounds for $t_\text{\tiny ES}$ for the OU and Wiener process
respectively, obtained from Eq.~\ref{eq:tes_lower_bound}.  Right: common
environment.} \label{fig:t_esd_vs_neg}
\end{figure}

In Fig.~\ref{fig:t_esd_vs_neg} we show $t_\text{\tiny ES}$ versus the initial
negativity $N_0$ for randomly generated Bell-state mixtures for the OU
process and the fGn with $H= \frac 12$. We can see that $t_\text{\tiny
ES}$ is bounded from below by a monotonically increasing function of
negativity, which itself diverges for $N_0 \rightarrow 1$, i.e. as 
the initial state gets closer to a pure Bell state.  
The analytical expression of this function is obtained by 
considering initial states belonging to a face of the 
Bell-state tetrahedron, and thus easily follows from
Eq. (\ref{eq:tstar_lower_bound}) by substituting $r=0$.
We have
\begin{equation}\label{eq:tes_lower_bound}
\beta(t_{\text{\tiny ES}}) = \frac{1}{4A} \log 
\left(\frac{1+N_0}{1-N_0}\right)
\end{equation}
where $A = 1$ for the independent-environments case and $A = 2$ for the
common-environment case.  Survival time is thus longer for larger values
of the  initial entanglement.  In the case of independent environments
the lower bound is larger than in the case of a common
environment, confirming the tendency of entanglement to be more robust
in the case of independent noises affecting the two qubits.  As opposed
to the entanglement-preserving time, the behavior of $t_\text{\tiny
ES}$ is comparable for the two considered processes. 
%%%
\section{Conclusions}
\label{sec:conclusions}
The decoherence caused by the interaction of a quantum system with the
external environment is one of the main obstacle 
to the large scale deployment of
quantum communication protocols and quantum information
processing. A deep understanding of the decoherence mechanisms 
and the ability to engineer the environment are thus in order to
obtain more robust quantum correlations and to design
robust implementations of quantum technologies. 
\par
In this paper, we have addressed the dynamics of a two-qubit system
interacting with classical noise generated by a stochastic Gaussian
process and leading to a dephasing time evolution.  In particular, we
considered two diffusion processes: the Ornstein-Uhlenbeck process,
characterized by a decoherence time $\gamma^{-1}$ and the fractional
Gaussian noise, characterized by the Hurst parameter $H$.  We computed
the time evolved density matrix of the two-qubit system by performing
the average over the stochastic processes, both in the case of
independent and common environments.  We have characterized the
trajectories of the system  inside the set of mixtures of Bell-states
and shown the occurence of sudden death of entanglement for certain sets
of initial quantum states.
\par
We introduced the entanglement-preserving time $t^*$ and the
entanglement-survival time $t_{\text{\tiny ES}}$ in order to analyze the
effects of the nature of noise on the decoherence mechanism. We found that
$t^*$ is larger for fGn than OU process and that a larger 
initial entanglement corresponds to a longer preserving time. 
We also found that $t^*$ is bounded from below by an increasing 
function of 
the initial negativity and that independent environments degrade
quantum correlations more weakly than a common one. 
Also the survival time $t_{\text{\tiny ES}}$ is bounded from below 
by a (different) increasing function of the initial negativity but, 
contrarily 
to the preserving time, has comparable values for 
the two considered processes.
\par
Overall, our results indicate that engineering the environment has only
a slight influence over the entanglement-survival time, i.e. the occurence
of entanglement sudden-death, while it represents a valuable resource 
to increase the entanglement-preserving time, i.e. to maintain entanglement 
closer to the initial level for a longer interaction time.
%%%%%%%%
\section*{Acknowledgments}
This work has been supported by the MIUR project 
FIRB-LiCHIS-RBFR10YQ3H.  
%%%%%%%%


\begin{thebibliography}{10}
\bibitem{helm09} J. Helm and W. T. Strunz, {\em Phys. Rev. 
A} {\bf 80} (2009) 042108.
\bibitem{helm11} J. Helm, W. T. Strunz, S. Rietzler, 
and L. E. W\"{u}rflinger, {\em Phys. Rev. A} {\bf 83} (2011) 042103.
\bibitem{crow2014classical} D.~Crow and R.~Joynt, {\em Phys. Rev. A }
{\bf 89} (2014)  042123.
\bibitem{wayne13} W. M. Witzel, K. Young, 
S. Das Sarma,  arXiv:1307.2597v1.
\bibitem{yu10} T.~Yu and H.~H. Eberly, {\em Opt. Commun.} 
{\bf 283}  (2010) 676.
\bibitem{arr10}
B. Bellomo, G. Compagno, A. D'Arrigo, G. Falci, R. Lo Franco, and E.
Paladino, Phys. Rev. A 81, 062309 (2010)
\bibitem{li11} J.-Q. Li and J.-Q. Liang, {\em Phys. Lett. A }
{\bf 375} (2011) 1496.
\bibitem{benedetti12} C. Benedetti, F. Buscemi, 
P. Bordone, M. G. A. Paris, 
\emph{Int. J. Quantum Inf.} {\bf 10} (2012) 1241005. 
\bibitem{bordone12} P. Bordone, F. Buscemi, C. Benedetti, 
\emph{Fluct. Noise Lett.} {\bf 11} (2012) 1242003.
\bibitem{rlf12} R. Lo Franco, B. Bellomo, E. Andersson, and G. Compagno, 
Phys. Rev. A {\bf 85} (2012) 032318
\bibitem{rlf13} R. Lo Franco, B. Bellomo, S. Maniscalco, and G. Compagno, 
Int. J. Mod. Phys. B {\bf 27} (2012) 1345053 
\bibitem{xu13}
J.-S. Xu, K. Sun, C.-F. Li, X.-Y. Xu, G.-C. Guo, E. Andersson, R. Lo
Franco, Nat. Commun. 4 (2013) 2851.
\bibitem{arr14}
A. D'Arrigo, R. Lo Franco, G. Benenti, E. Paladino, and G. Falci, 
Ann. Phys. {\bf 350}, (2014) 211.
\bibitem{paraoanu14} J. Li, M. P. Silvestri, K. S. Kumar, 
J.-M. Pirkkalainen,A. Veps\"al\"ainen, W. C. Chien, J. Tuorila, M. A. 
Sillanp\"a\"a, P. J. Hakonen, E. V. Thuneberg, and G. S. 
Paraoanu, \emph{Nat. Commun.} {\bf 4} (2013) 1420.
\bibitem{meno}
K. Kakuyanagi, T. Meno, S. Saito, H. Nakano, K. Semba, H.
Takayanagi, F. Deppe, 
A. Shnirman, \newblock \emph{Phys. Rev. Lett.} {\bf 98} (2007) 047004
\bibitem{tsai}
F.~Yoshihara, K.~Harrabi, A.~O. Niskanen, Y.~Nakamura, and J.~S. Tsai,
\newblock \emph{Phys. Rev. Lett.} {\bf 97} (2006) 167001
\bibitem{pal0x}
E.~Paladino, L.~Faoro, G.~Falci, and R.~Fazio,
\emph{ Phys. Rev. Lett.} {\bf 88} (2002) 228304; 
 G.~Falci, A.~D'Arrigo, A.~Mastellone, and E.~Paladino,
 \emph{Phys. Rev. Lett.} {\bf 94} (2005) 167002
\bibitem{bukard} G. Bukard, \emph{Phys. Rev. B} {\bf 79} 125317 (2009).
\bibitem{bergli12}H. J. Wold, H. Brox, Y. M. Galperin, and 
J. Bergli \emph{Phys. Rev. B} {\bf 86}  (2012) 205404.
\bibitem{pal12} R. Lo Franco, A. D'Arrigo, G. Falci, G. Compagno, 
and E. Paladino, Phys. Scripta {\bf T147} (2012) 014019
\bibitem{nonmark} C. Benedetti, M. G. A. Paris, and 
S. Maniscalco, \emph{Phys. Rev. A} {\bf 89} (2014) 012114.
\bibitem{benedetti2013dynamics} C.~Benedetti, F.~Buscemi, 
P.~Bordone and M.~G.~A. Paris, {\em Phys. Rev. A}   {\bf 87}  (2013)   052328.
\bibitem{mannone12} M. Mannone, R. Lo Franco and 
G. Compagno, \emph{Phys. Scr.} {\bf T153} (2013) 014047.
\bibitem{rev14} E.~Paladino, M.~Galperin, 
Y., G.~Falci and B.~L. Altshuler, {\em Rev. Mod.  Phys.} {\bf 86}  (2014) 361.
\bibitem{paris14} C. Benedetti, M. G. A. Paris, \emph{Int. 
J. Quantum Inform.} {\bf 12} (2014) 1461004.
\bibitem{benedetti14} C. Benedetti, M. G. A. Paris, 
\emph{Phys. Lett. A} {\bf 378} (2014) 2495.
\bibitem{parisF14} M. G. A. Paris, \emph{Physica A} {\bf 413} (2014) 256.
\bibitem{bergli09} J. Bergli, Y. M. Galperin and B. L. 
Altshuler, \emph{New J. Phys.} {\bf 11} (2009) 0250022.
\bibitem{astafiev2004quantum} O.~Astafiev, Y.~A. Pashkin, 
Y.~Nakamura, T.~Yamamoto and J.~S. Tsai, \emph{Phys.   
Rev. Lett.} {\bf 93}  (2004)   267007.
\bibitem{bergli06} Y. M. Galperin, B. L. Altshuler, 
J. Bergli and D. V. Shantsev, \emph{Phys. Rev. Lett.}, {\bf 96} (2006) 097009.
\bibitem{abel08} B. Abel and F. Marquardt, \emph{Phys. Rev. 
B} {\bf 78} (2008) 201302(R)
\bibitem{averin04} K. Rabenstein, V. A. Sverdlov and D. V. Averin,
{\it JETP Lett. } {\bf 79} (2004) 646.
\bibitem{shnirma} Y. Makhlin and A. Shnirma,{\it Phys. Rev. Lett.} {\bf  92} (2004) 178301.
\bibitem{shibata} M. Bana, S. Kitajimac, F. Shibata, 
{\it Phys. Lett. A} {\bf 349} (2006) 415.
\bibitem{sarma08} \L. Cywi\`nski, R. M. Lutchyn, C. P. Nave, S. Das Sarma, 
{\em Phys. Rev. B} {\bf 77} (2008) 174509.
\bibitem{sarma13b}
J.-T. Hung, L. Cywi\`nski, X. Hu, S. Das Sarma, 
Phys. Rev. B {\bf 88} (2013) 085314.
\bibitem{yu2006sudden} T.~Yu and J.~H. Eberly, 
{\em Opt. Comm.} {\bf 264}  (2006) 393.
\bibitem{yu2009sudden} T.~Yu and J.~H. Eberly, 
{\em Science} {\bf 323}  (2009) 598.
\bibitem{palma03} G. De Chiara and G. M. Palma, 
\emph{Phys. Rev. Lett.} {\bf 91} (2003) 090404.
\bibitem{spagnolo09} A. Flasconaro, B. Spagnolo, 
\emph{Phys. Rev. E} {\bf 80} (2009) 041110.
\bibitem{mandelbrot1968fractional} B.~B. Mandelbrot 
and J.~W. Van~Ness, {\em SIAM Rev.} {\bf 10} (1968) 422.
\bibitem{puri2001mathematical} R.~R. Puri, {\em Mathematical 
methods of quantum optics} (Springer, Berlin, 2001).
\end{thebibliography}
\end{document}